\documentstyle[multicol,aps,prl,epsf]{revtex}

\begin{document}
\pagestyle{empty}

\begin{multicols}{2}
\narrowtext
\parskip=0cm

\noindent
{
\bf 
Comment on ``Ising Spin Glasses in a Magnetic Field''}

In a very interesting paper \cite{HOUMAR} Houdayer and Martin analyze
the $T=0$ $3d$ EA spin glass with a magnetic field $B$. By using a
new, powerful method, they determine an effective critical field $B_c$
(given by the crossings of their function $r$, the rate at which
ground states change) as a function of the lattice size $L$.  They use
their results to deduce that the model is behaving like in the droplet
approach and not like the mean-field theory, that implies the survival
of a phase transition for finite field $B$. We show here, by using
some unpublished data \cite{UNPUBL}, that the very interesting method
and numerical results of \cite{HOUMAR} are completely compatible with
the behavior implied by the Replica Symmetry Breaking (RSB) theory.

In \cite{MAPAZU} we have introduced a dynamical approach that has
allowed us to determine the value of the minimal allowed overlap,
$q_{min}$, for the $4d$ EA model in a non-zero magnetic field.  By
establishing that this value is different from $q_{ave}$, the average
overlap measured on medium size samples, we have shown that also in
$4d$ the phase transition to a RSB phase survives after applying a
magnetic field.  We use here analogous data for the $3d$ case
\cite{UNPUBL}, that also clearly show the survival of the
Almeida-Thouless line, to interpret the result of \cite{HOUMAR}.

\begin{figure}
\centerline{
\epsfysize=0.6\columnwidth{\epsfbox{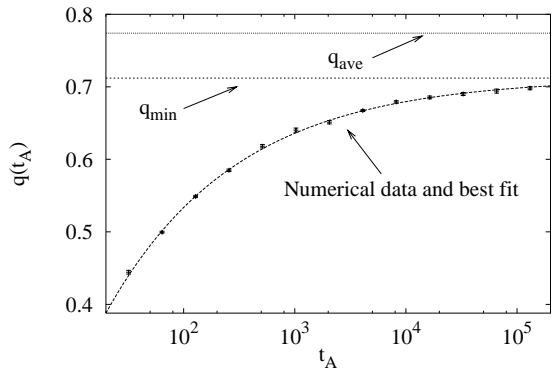}}}
\narrowtext{
\caption{$q(t_A)$ versus $t_A$ at $T=0.5$, $B=0.4$.  
\protect\label{fig1} } }
\end{figure}

\begin{figure}
\centerline{
\epsfysize=0.6\columnwidth{\epsfbox{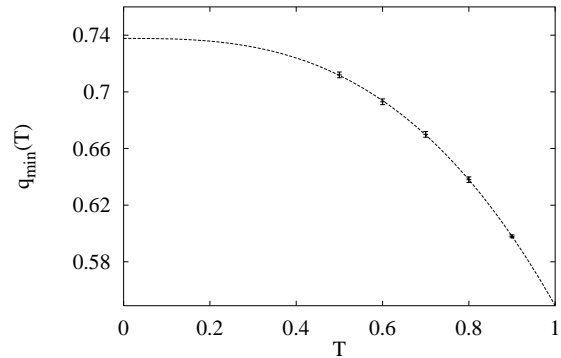}}}
\narrowtext{
\caption{ 
$q_{min}(T)$ and best power fit.
\protect\label{fig2}
}  
}
\end{figure}

In our approach we use a large lattice (for example $L=32$), and we
select an annealing schedule to run a Metropolis dynamics.  We select
a set of decreasing values of the temperature $\{T_n\}$, starting from
high $T\gg T_c$, and we spend $n\cdot 2^k$ updating steps at each
$T_n$ value.  We compute $q(t_A)$, where $t_A\equiv 2^k$, for
increasing $k$ values.  The limit for $t_A\to\infty$ of $q(t_A)$ is
just $q_{min}$.  The value of $q_{ave}$ is computed with usual
equilibrium tempering simulations on smaller lattices.  We show in
figure (\ref{fig1}) what happens in $3d$ at $B=0.4$ and $T=0.5$, where
RSB is clear: the fit of the data is very precise, and gives a value
of $q_{min}$ that clearly differs from $q_{ave}$.  Ref. \cite{HOUMAR}
deals with $T=0$, so in figure (\ref{fig2}) we extrapolate
$q_{min}(T)$ at $B=0.4$ down to $T=0$.  A simple power fit is very
good, giving $q_{min}(T=0)=0.74$.  In terms of the percentage of
different spins $x$ of \cite{HOUMAR} this means that at $B=0.4$ we
find $x\simeq 0.13$, corresponding to a jump with overlap $0.74$: in
the conditions of \cite{HOUMAR} (where $x=0.15$) one expects as
$L\to\infty$ a crossing of the $r$ curves close to $B_c=0.4$.

This is well compatible with the data of \cite{HOUMAR}: their crossing
point moves from $B\simeq 2.1$ to $B\simeq 1.2$ when $L$ goes from $3$
to $8$: in these conditions it is difficult to discriminate between
the droplet $B_c=0$ and the $B_c\simeq 0.4$ that follows from RSB. RSB
predicts that $r \simeq L^{\frac32}$ at small $B$, which is not far
from the observed behavior.  A main feature of short range models is
that one needs very large $L$ values to determine the asymptotic value
of $q_{min}$: the value of $q_{min}$ becomes sizably different from
zero already for a small $B$ ($q_{min} \approx B^{0.3}$ in $4d$
against $q_{min} \approx B^{\frac23}$ in $d=\infty$).  This implies
that the width at $L=\infty$ of $P(q)$ becomes much smaller as soon as
$B$ is switched on, and finite volume effects become much more
important.  This effect is exacerbated by the presence of the very
large, low $q$ tail of $P(q)$ at $L$ not too large (see \cite{MANAZU}
for the analogous $4d$ case), that disappears very slowly for
increasing $L$.  The new method of \cite{HOUMAR} is very promising,
and going to larger lattices will probably allow one further
verification of RSB for finite $d$ spin glasses.

These simulations have been run at ZIB-Berlin.  We thank H. Stueben
for his crucial and extensive help: without him this work would have
been impossible.  We acknowledge W. Janke and D. Johnston contribution
in setting up this project.

\noindent 
E. Marinari, G. Parisi and F. Zuliani


%

\end{multicols}
\end{document}